\documentclass[%
 reprint,
 amsmath,amssymb,
 aps,
prx]{revtex4-2}

\usepackage{graphicx}% Include figure files
\usepackage{dcolumn}% Align table columns on decimal point
\usepackage{bm}% bold math
\usepackage{hyperref}
\usepackage{lipsum}
\usepackage{enumerate}  

\begin{document}

\title{Experimental Demonstration of Dephasing Reduction in an Optically Guided Laser-Plasma Accelerator}

\author{Ronan~Lahaye}
\author{Igor~A.~Andriyash}
\author{Julien~Gautier}
\author{Olena~Kononenko}
\author{Adrien~Leblanc}
\author{Jean-Philippe~Goddet}
\author{Amar~Tafzi}
\author{Cedric~Thaury}
 \email{cedric.thaury@ensta.fr}
\affiliation{Laboratoire d’Optique Appliqu\'ee, ENSTA, CNRS, Ecole polytechnique, Institut Polytechnique de Paris, 828 Bd des Mar\'echaux, 91762 Palaiseau, France}

\begin{abstract}
Laser-plasma accelerators offer a compact means of producing high-energy electron beams, but their performance is fundamentally limited by dephasing between the accelerated electrons and the plasma wave. To overcome this limitation, we investigate the combination of plasma density tapering and optical guiding to extend the effective acceleration length. Using a Joule-class femtosecond laser coupled into an optical-field-ionized plasma waveguide with a controlled density gradient, we experimentally achieve electron beam energies exceeding 1.6 GeV, a 40\% increase compared to the constant-density case. Particle-in-cell simulations reproduce the main experimental features and reveal the central roles of delayed injection, nonlinear laser evolution, and self-focusing in enhancing energy gain. 
\end{abstract}

\maketitle

\section{Introduction}

Laser-plasma acceleration is a promising alternative to conventional accelerators due to the high-amplitude accelerating fields generated when an ultra-intense laser pulse travels through an underdense plasma, driving a strong plasma wave in its wake~\cite{Fainberg1960,PhysRevLett.43.267}. These accelerating fields, known as wakefields, can reach up to 100 GV/m. Maximizing the interaction length between the accelerated electron bunch and the accelerating field is essential to fully exploit these fields. However, this length is limited by both laser diffraction in the plasma and dephasing between the electron bunch and the driving pulse.

Diffraction causes the laser beam to diverge after propagating a certain distance in the plasma, reducing its intensity to a point where it can no longer efficiently excite a plasma wave. This limitation can be overcome by using a plasma waveguide, which consists of a plasma with a curved radial density profile that helps to maintain the laser's focus over extended distances~\cite{1993PhRvL..71.2409D}. 

Dephasing arises from the velocity mismatch between the electron beam, which travels at $v_e\approx c(1-1/2\gamma^2)\approx c$, and the laser, which propagates in the plasma at the group velocity $v_g\approx(1-n_e/2n_c)<c$, where $n_e$ and $n_c$ are the electron and critical plasma densities, respectively. This mismatch causes the electron beam to drift from the back toward the front of the plasma cavity, gradually experiencing a weaker accelerating field. In extreme cases, the electron bunch may even move beyond the cavity center, thus experiencing a decelerating field. The characteristic length of this process, or the so-called dephasing length $L_d$, is proportional to $n_e^{-3/2}$~\cite{RevModPhys.81.1229}. Thus, dephasing can be mitigated by reducing the plasma density $n_e$, but at the cost of a lower accelerating field and reduced self-focusing.

To ensure that the electrons remain in the accelerating phase, the length of the plasma cavity can be dynamically shortened ensuring that the electron beam remains at the back of the cavity where the electric field is maximum~\cite{PhysRevE.63.056405}. This technique, known as tapering, has been demonstrated in pioneering experiments using either a linear density gradient~\cite{2011JKPS...59.3166J,10.1063/1.4807440,2019NatSR...911249A, 2024PhyS...99a5603N} or a density step~\cite{2015PhRvL.115o5002G}. Yet, these experiments have so far yielded electron beams with energies below 500 MeV, primarily because of the lack of efficient laser guiding.
% Dephasing can be more effectively mitigated by dynamically shortening the length of the plasma cavity as the electron beam approaches the laser pulse, ensuring that it remains at the back of the cavity where the electric field is maximum~\cite{PhysRevE.63.056405}. 

Here, we show that this approach can be combined with optical guiding to counteract diffraction and mitigate dephasing simultaneously. In the first section, we use a simple analytical model to examine the impact of a rising density gradient on dephasing. We then present experimental results demonstrating the production of electron beams up to 1.8 GeV with a joule-class laser, followed by an analysis using particle-in-cell (PIC) simulations.

%In this study, we propose to address dephasing by tailoring the phase velocity of the wake by shaping the longitudinal density profile along the propagation, combined with optical guiding to mitigate diffraction and increase the acceleration length.

\section{Dephasing in a rising density gradient}
\label{sec:theo}
In this section, we use a simple model to demonstrate how longitudinal shaping of the plasma density can reduce the effect of dephasing. To mitigate dephasing, we ideally want the electron bunch to remain in the same phase of the wakefield throughout the entire propagation, that is, the phase velocity of the wakefield at the position of the electron bunch should be equal to the beam velocity $v_e\approx c$. The wakefield phase can be expressed as:
\begin{equation}
    %\phi(z,t)=\frac{\pi}{r_B(z_L(t))}(z_L(t)-z),
    \phi(z,t)=\pi\; \cfrac{z_L(t)-z}{r_B(z_L)},
    \label{eq:phase_general}
\end{equation}
where $r_B\propto 1/\sqrt{n_e}$ is the radius of the accelerating cavity, and $z_L$ is the position of the laser front, which we assimilate with the upper boundary of the first plasma cavity. 
% Supposing $\lambda_p\ll n_e/(dn_e/dz)$ and $(z_L-z)\ll n_e/(dn_e/dz)$, we have [ref Bulanov+RMP]

\begin{figure*}[t!]
    \centering
    \includegraphics[width=.9\textwidth]{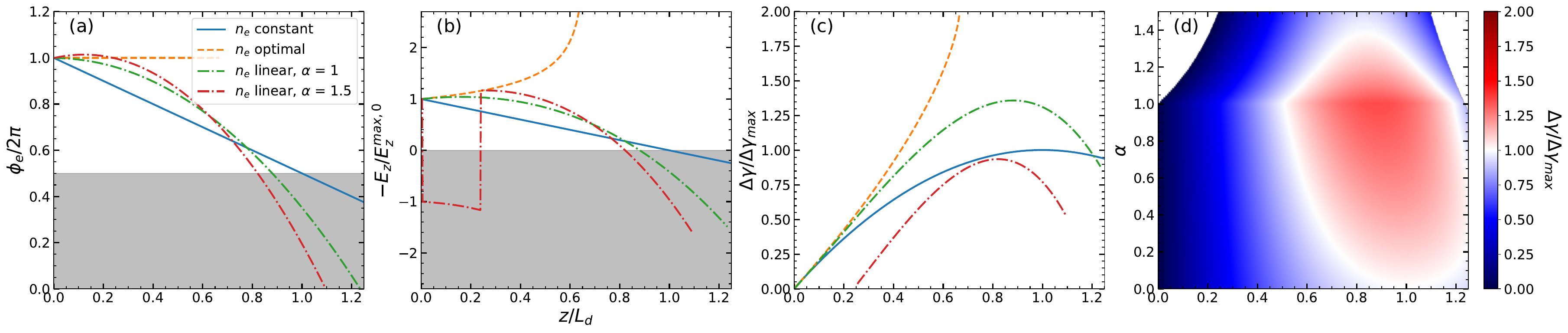}
    \caption{Comparison between the constant density case, the density profile defined by Eq.~(\ref{eq:prof_ideal}), and two linear density profiles given by Eq.~(\ref{eq:prof_lin_Ld}) for $\alpha=1$ and $\alpha=1.5$. Panel (a) shows the phase of an electron bunch injected at the back of the cavity at $z_0=0$, panel (b)  displays the longitudinal electric field experienced by the electron bunch. Panel (c) shows the energy gain as a function of $z$. Panel (d) shows the energy gain as a function of $z$ and $\alpha$.%On panel~(b), the electric field for the case $\alpha=1.5$ changes sign and is decelerating at the beginning of the propagation, when the phase is greater than $2\pi$, corresponding to an electron bunch crossing the lower boundary of the cavity and experiencing the decelerating field in the front of the second cavity. When the phase becomes lower than $2\pi$, the electron bunch can be injectedat the back of the first cavity and experiences an accelerating field.
    The gray area indicates the region where the field is decelerating.
    On panels~(b),~(c) and ~(d), the electric field and energy gain computed for the linear cases  are truncated at the $z$ position where $\phi_e=0$, which corresponds to the electron bunch reaching the position of the laser pulse.}
    \label{fig:comp_gain}
\end{figure*}

From this, the phase velocity of the wakefield is given by
\begin{equation}
    v_{\phi}(\xi,z)= \left(\frac{d\phi}{dt}\right)  \left(\frac{d\phi}{dz}\right)^{-1}\!\! \approx v_g(z)\left(1+\frac{\xi}{2n_e}\frac{dn_e}{dz}\right),
    \label{eq:v_phi_general}
\end{equation}
where $\xi=z_L-z$ represents the distance to the front of the laser~\cite{PhysRevE.58.R5257,RevModPhys.81.1229}. This expression assumes that the density gradient scale length, $n_e/(dn_e/dz)$, is much larger than both the plasma wavelength $\lambda_p$ and $\xi$.
These assumptions are well verified in most practical cases. %On the other hand, 
The equation also assumes that $r_B$ depends only on the local density. This hypothesis is more problematic, as in reality the length of the cavity also depends on the laser intensity, which varies during propagation. Therefore, Eq.~(\ref{eq:v_phi_general}) should be regarded more as a qualitative analysis tool, than as a means of prediction.
% and $v_g$ the group velocity of the laser pulse in the plasma. The group velocity can be expressed as
%\begin{equation}
%    \frac{v_g}{c}=1-\kappa\frac{n_e}{n_c},
%    \label{eq:v_g_ne}
%\end{equation}
%with $n_c$ the critical density and $\kappa$ a numerical factor, which takes different values depending on the regime of acceleration. In the linear regime, $\kappa\sim0.5$, while in the blow-out regime,  $\kappa\sim1.5$. Under these assumptions, the velocity of the back of the first cavity can be computed by evaluating Equation~\ref{eq:v_phi_general} for $\xi=2r_B$, which yields
%\begin{equation}
%    v_B=c\left(1-\kappa\frac{n_e}{n_c}+\left(1-\kappa\frac{n_e}{n_c}\right)\left(\frac{n_e}{n_0}\right)^{-1/2}\frac{r_{B,0}}{n_e}\frac{dn_e}{dz}\right).
%    \label{eq:v_back_general}
%\end{equation}
% Moreover, the effect of the gradient is twice as significant in the second cavity where $\xi=4r_B$. 
Specifically, the equation shows that for an increasing density profile: 
\begin{enumerate}[(i)]
    \item The phase velocity $v_\phi$ exceeds the laser group velocity $v_g$.
    \item The velocity difference $v_\phi-v_g$ is proportional to the distance to the laser $\xi$.
    \item The phase velocity $v_\phi$ can become superluminal if the plasma density increases too steeply.
\end{enumerate}

The condition that the phase velocity at the rear of the bubble is equal to $c$, \emph{i.e.} $v_\phi(2r_b,z)=c $, defines a differential equation  for the plasma density  $n_e(z)$ that can be easily solved, leading to~\cite{PhysRevE.77.025401,10.1063/1.4946018}
\begin{equation}
    n_e(z)=\frac{n_0}{\left(1-z/L_0\right)^{2/3}},
    \label{eq:prof_ideal}
\end{equation}
with $n_0=n_e(0)$, $L_0=(2/3)L_{d0}$, and $L_{d0}$ the dephasing length at $n_0$.  Such a density profile would be challenging to achieve experimentally due to the limited means for shaping the density. Moreover, it was derived under the assumption of constant laser intensity, which is known to be inaccurate; as a result, the actual profile that would maintain the beam at the rear of the cavity could differ significantly from Eq.~(\ref{eq:prof_ideal})~\cite{StreeterEAAc}. Therefore, it is meaningful to analyze the dynamics of dephasing in simpler density profiles that are easier to achieve. In the following, we will consider linear density profiles which are easier to create experimentally.

In the general case, where the wakefield velocity is not locked to the electron velocity, the co-moving position of the bunch is  $\xi_e=z_L-z_e$, and the phase of the electron bunch in the cavity, assuming $v_e=c$ is
\begin{equation}
    \phi_e(z_e)=\frac{\pi}{r_B(z_e)}\left(\int_{z_0}^{z_e}\left(\frac{v_g(z)}{c}-1\right)dz +\xi_0\right),
\end{equation}
where $z_0$ is the injection position, $\xi_0=\xi_e(z_0)$, and both $r_B$ and $v_g$ are functions of the local plasma density $n_e(z)$.  Due to the velocity mismatch, the electron beam experiences a varying electric field during acceleration. In the bubble regime, this field is given by :
\begin{equation}
    E_z(z_e)=-\frac{e n_e(z_e)}{\epsilon_0}\left(\xi_e(z_e)-r_B(z_e)\right)\mathrm{,}
    \label{eq:Ez}
\end{equation}
with $e$ the electron charge and $\epsilon_0$ the vacuum permittivity \cite{PhysRevSTAB.10.061301}. The energy gain from the injection position is then directly obtained from Eq.~(\ref{eq:Ez}):
\begin{equation}
    \Delta\gamma(z_e)=\frac{q}{mc^2}\int_{z_0}^{z}E_z(z')dz'\mathrm{.}
\end{equation}

%Let us now consider an electron bunch inside the cavity at the co-moving position $\xi_e=z_L-z_e$, assuming that $v_e=c$.   The longitudinal electric field experienced by the bunch is

%The phase of the electron bunch in the cavity is
% Using Eq.~\ref{eq:v_g_ne}, we can then write
%\begin{equation}
%    \phi_e(z)=\frac{\pi}{r_B(z)}\left(\int_{z_i}^z\left(-\kappa\frac{n_e}{n_c}\right)dz'+\delta_0\right).
%\end{equation}
%This profile is represented in Figure\ref{fig:ne_ideal_vs_lin}.
%Computing the energy gain in the case of such a profile leads to $\Delta\gamma(L_0)=2\Delta\gamma_{max}$, with $\Delta\gamma_{max}$ the maximum energy gain achieved at dephasing for $n_e=n_0$.
%This density profile is challenging to reproduce experimentally,

 %so we will consider the case of a linear density profile of the form

The three quantities $\phi_e$, $E_z$ and $\Delta\gamma$ are plotted in Fig.~\ref{fig:comp_gain}(a-c) for a constant density $n_0$, the profile given by  Eq.~\ref{eq:prof_ideal}, and two linear density profiles defined as 
\begin{equation}
    n_e(z)=n_0\left(1+\alpha\frac{z}{L_d}\right),
    \label{eq:prof_lin_Ld}
\end{equation}
with $\alpha =1$ (green curve) and $\alpha = 1.5$ (red curve).

In the constant-density case, the dephasing rate is constant (\emph{i.e.}, $\phi_e$ is a linear function of $z$), and the beam experiences progressively lower fields as it drifts through the cavity. Consequently, the energy gain follows a parabolic curve, reaching a maximum of $\Delta \gamma_{max}$ at $L_{d0}$, when the beam reaches the center of the cavity.
For the optimal density profile, the beam remains in the same phase by design, while the amplitude of the electric field experienced by the beam increases during acceleration as $E_z \propto \sqrt{n_e}$. Consequently, the energy gain increases much more steeply than in the reference case, reaching its maximum at $z = L_0$~\cite{PhysRevE.77.025401,10.1063/1.4946018}, where the cavity becomes infinitely small, stopping further acceleration. 

For a linear profile with $\alpha = 1$,  $\phi_e$ follows the optimal phase over a short distance (approximately $0.05 L_{d}$), after which it dephases more rapidly than in the constant-density case, because the group velocity decreases with increasing plasma density. The longitudinal electric field experienced by the electron bunch follows a similar trend, initially increasing in amplitude for small $z$, then decreasing faster than in the constant-density case as the beam approaches the center of the cavity. Overall, as the beam spends more time in the region of high fields, this results in an energy gain that exceeds $\Delta \gamma_{\text{max}}$ for $z \leq L_{d0}$, with the maximum energy gain reached around $z \sim 0.88 L_{d0}$, where $\Delta \gamma \approx 1.36 \Delta \gamma_{\text{max}}$.

For a linear profile with $\alpha=1.5$, $\phi_e$ initially increases, which means that the wakefield is superluminal. Such superluminal propagation should prevent any electron injection.  We therefore consider that the injection occurs at $z\approx0.2L_d$, where $d\phi_e/dz$ changes sign, meaning that the wakefield propagation becomes subluminal.%, and the electric field becomes accelerating, as seen in Figure~\ref{fig:comp_gain}~(b). 
The evolution of the electric field has the same shape than in the case $\alpha=1$, but the dephasing rate occurs faster, which results in a maximum energy gain reached around $z\approx0.94L_d$ where $\Delta\gamma\approx0.94\Delta\gamma_{\text{max}}$. 

% The acceleration dynamics is then quite similar to that in case $\alpha=1$, except that a longer plasma length is required to reach the same energy due to the delayed injection. In that case, the maximum energy gain is reached around $z\approx0.94L_d$ where $\Delta\gamma\approx0.94\Delta\gamma_{\text{max}}$.

In conclusion, the linear profile with $\alpha=1$ stands out as the most promising among the linear density profiles, as illustrated in Fig.~\ref{fig:comp_gain}(d). We observe that for $\alpha>1$, the initial superluminal phase velocity of the wake prevents injection, and results in a shorter effective acceleration length. For $\alpha<1$, the mitigation of dephasing is less effective and results in a lower maximum energy gain than for $\alpha=1$, which is the optimum as it minimizes dephasing while maintaining a phase velocity that remains consistently subluminal, ensuring a higher maximal energy gain.

\section{Experimental results}
%We set up an experiment to test different linear profiles and demonstrate how they can increase the energy of accelerated electrons.
We set up an experiment to test this conclusion and determine the optimal linear-density gradient for maximizing the electron beam energy. In order to observe only effects related to dephasing, it is necessary to ensure that the beam acceleration is not limited by another phenomenon. A plasma waveguide is thus used to overcome laser diffraction and maintain the laser's focus throughout its propagation in the plasma~\cite{1993PhRvL..71.2409D}. %Additionally, we operate the accelerator at plasma densities where the depletion length~\cite{PhysRevSTAB.10.061301} ($\sim n_e/n_c c\tau$, with $\tau$ the laser pulse duration) exceeds the target length.

The experiment was conducted at LOA using the 'Salle Jaune' laser system. %The setup is shown in Fig.~\ref{fig:schema_SJ}.
The plasma waveguide was produced through hydrodynamic optical-field-ionization (HOFI). This method involves focusing a laser along a line longer than the gas target to ionize it and create a plasma filament. The hydrodynamic expansion of this filament during a few nanoseconds, combined with the ionization of the remaining neutrals, results in the formation of a waveguide capable of maintaining the laser pulse focused over long distances~\cite{PhysRevE.97.053203,Smartsev:19}.
Here, the guiding laser ($E\sim2.3$~mJ, $\tau\sim30$~fs) was focused %into the nozzle 
by an off-axis axiparabola~\cite{Oubrerie_2022} with a nominal focal length $f_0 = 200$~mm and maximum focal depth $\delta_0=30$~mm. The focal line of this axiparabola is defined by $f(r) = f_0 + 1/a\ln (r/R\times\text{e}^{a\delta_0})$, with $a=1/\delta_0\ln(R/r_{\text{hole}})$, where $r$, $r_{\text{hole}}=6~\text{mm}$ and $R=38.1~\text{mm}$ denote the radial coordinate, the radius of the hole and the radius of the axiparabola, respectively. A deformable mirror was used to optimise the focal spot of the beam focused by the axiparabola. %The expansion of the filament over $\sim2$~ns leads to the formation of a waveguide in which the drive laser ($E\sim1$~J, $\tau\sim30$~fs) can be guided.
%During the propagation, the driver beam  generates a wakefield in which electrons are accelerated over a long distance and is not limited by diffraction. 

The driver beam had a duration of 30 fs. It was focused at the entrance of the waveguide by an F/18 spherical mirror used off-axis, with the astigmatism introduced by the off-axis geometry corrected by a second deformable mirror. The focal spot had an FWHM diameter of $D_{\text{FWHM}}=26$~$\mu$m, and the energy encircled within the first zero ring reached $1.05\pm~0.05$ J, corresponding to 73\% of the total laser energy. The laser pointing stability was improved using an active stabilization system~\cite{10.1063/1.3556438}, achieving a peak-to-peak stability of $0.7 D_{1/2}$, and standard deviations $\sigma_x=D_{\text{FWHM}}/5.7$, $\sigma_y=D_{\text{FWHM}}/6.3$ in the horizontal and vertical directions, respectively. This level of stability is sufficient to ensure substantial laser injection into the plasma waveguide on each shot, even though the fraction of coupled energy may still vary.

\begin{figure}
    \centering
    \includegraphics[width=\linewidth]{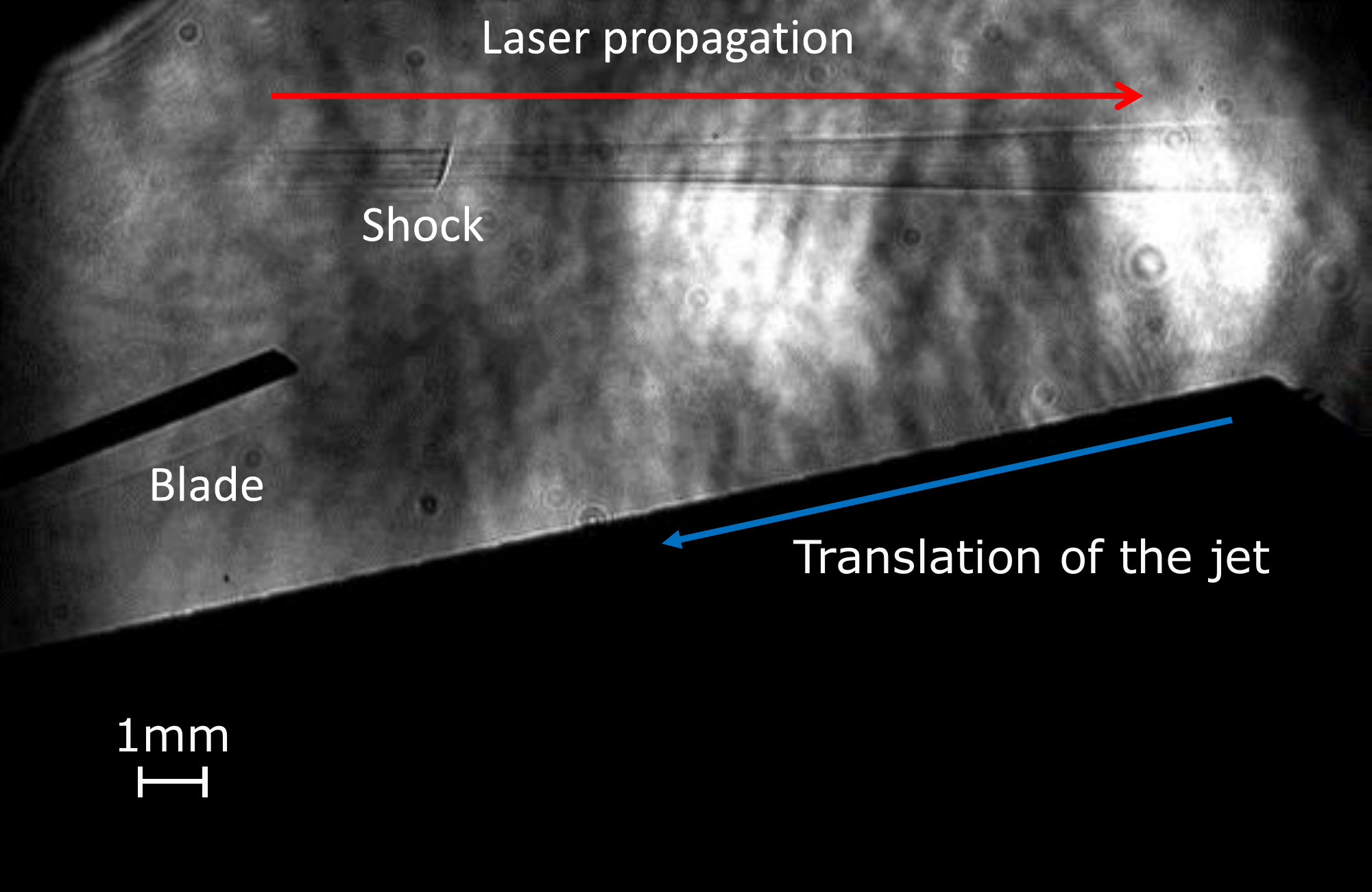}
    \caption{Shadowgram of the tilted target, showing the plasma created by the laser, the blade to trigger the injection and the shock created by the blade.}
    \label{fig:enter-label}
\end{figure}

To produce the linear density gradient, a 20~mm-long slit nozzle was mounted on a rotative stage, which allows us to %monitor and
change the angle between the laser axis and the gas nozzle. The density profiles measured for different angles are provided in the Supplementary Materials (SM).
The nozzle was fed with a gas mixture of dihydrogen and 1\% nitrogen. A blade made of a 200 $\mu$m-thick silicon wafer was used to partially obstruct the gas flow at the entrance of the gas jet to generate a hydrodynamic shock. The purpose of this shock is to produce a sharp density transition that can trigger the injection of electrons into the wakefield through shock-assisted ionization injection~\cite{2015NatSR...516310T}. The blade was also used to reduce the effective target length by adjusting the position of the nozzle underneath it, allowing us to maintain the relative positioning between the density transition and the focal plane of the main laser.
Finally,  a probe laser beam (a few~mJ, $\tau\sim30$~fs) is used to perform shadowgraphy measurements of the interaction and to measure the electronic density with a wavefront sensor (Phasics SID4-UHR).
An example of shadowgraphy is displayed in Fig.~\ref{fig:enter-label}. It shows  the position of the blade, the plasma shadow, the generated shock and the translation axis of the gas nozzle.
A simple electron spectrometer, consisting of a 40~cm-long, 0.85~T magnetic dipole and a scintillating screen, was used to disperse and detect the electron beam (see details in SM).

% \begin{figure}
%     \centering
%     \includegraphics[width=\linewidth]{Figures/schema_SJ.pdf}
%     \caption{Experimental setup}
%     \label{fig:schema_SJ}
% \end{figure}

% \begin{figure}
%     \centering
%     \includegraphics[width=0.5\linewidth]{Figures/spectrum_20240530_flat.pdf}
%     \caption{Case without gradient for different densities}
%     \label{fig:20240530_flat}
% \end{figure}

\begin{figure}
    \centering
    \includegraphics[width=\linewidth]{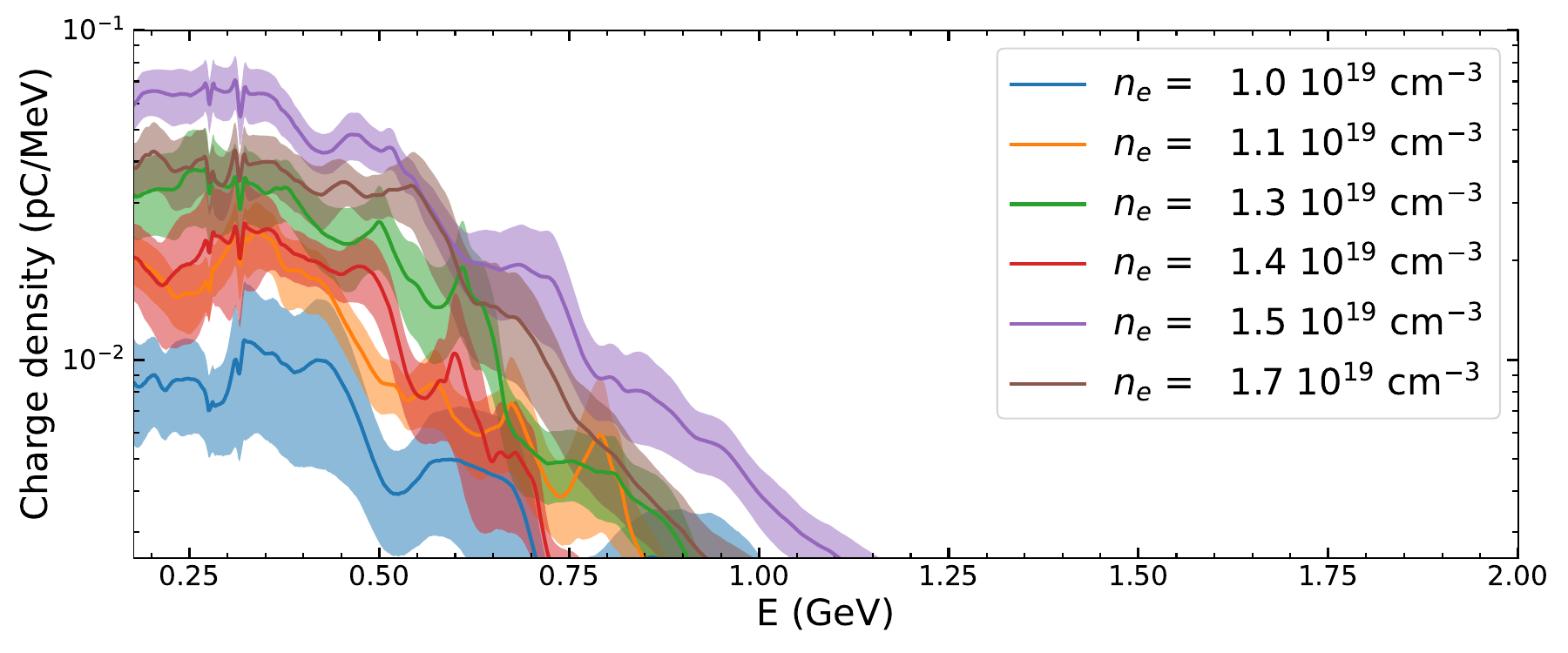}
    \caption{Mean spectra obtained  the absence of a gradient, i.e., with an untilted gas jet, for different operating pressures. The mean is computed over 10 shots, without excluding any, for each pressure. The shaded area corresponds to the standard deviation over the same 10 shots.}
    \label{fig:20240530_flat}
\end{figure}

We first produced electron beams using the full target length and without introducing a density gradient (see density profile in Supplementary Fig.~S3). The mean spectra obtained at various plasma densities are shown in Fig.~\ref{fig:20240530_flat}.  The electron spectra are continuous, indicating that trapping at the density transition is not sufficiently efficient to fully load the wakefield and suppress further injection via ionization injection~\cite{10.1063/1.2179194}. Since the spectrum obtained are continuous, we chose to compare the cut-off energy for different configurations. We defined the cut-off at a threshold of 2.5~fC/MeV, which corresponds to roughly three times our detection threshold. The cut-off energy ranges from 600~MeV to 1~GeV, with a maximum of  $1.1\pm0.06$~GeV, obtained for a backing pressure of 27 bar and 
a measured density without the guide of $n_e\sim1.5\times10^{19}~\text{cm}^{-3}$. %, in agreement with previous studies without density gradient~\cite{2022LSA....11..180O}. 
 At higher operating pressures, the cut-off energy decreases, which indicates that dephasing becomes significant. In the optimal case, the electron density in the guiding channel is estimated to be of the order of $3\times10^{18}~\text{cm}^{-  3}$\cite{PhysRevE.97.053203}. %At these densities, the laser is likely  depleted after $\approx 10-15$~mm of propagation, so only a portion of the plasma is effectively used for acceleration.

% of 800~MeV to 1~GeV, with no clear dependency with the density. When averaging over 10 shots, we obtain a maximum cut-off energy of $829\pm238$~MeV, for a density measured without the guide of $n_e\sim1.5\times10^{19}~\text{cm}^{-3}$. We can estimate the density in the guiding channel in this case to be on the order of $n_e\sim3\times10^{18}~\text{cm}^{-3}$ [ref /5].
% \iffalse
% \begin{figure}
%     \centering
%     \includegraphics[width=\linewidth]{Figures/dens_profile_flat_mirrored.pdf}
%     \caption{Density profile recorded without gradient.}
%     \label{fig:dens_flat}
% \end{figure}

% \begin{figure}
%     \centering
%     \includegraphics[width=\linewidth]{Figures/dens_profile_8_deg.pdf}
%     \caption{Density profile recorded with a tilting angle of 8°.}
%     \label{fig:dens_8deg}
% \end{figure}

% \begin{figure}
%     \centering
%     \includegraphics[width=\linewidth]{Figures/dens_profile_12_deg.pdf}
%     \caption{Density profile recorded with a tilting angle of 12°.}
%     \label{fig:dens_12deg}
% \end{figure}
% \fi
% \begin{figure}
%     \centering
%     \includegraphics[width=0.5\linewidth]{Figures/spectrum_20240606.pdf}
%     \caption{Case with gradient and several target lengths}
%     \label{fig:20240606_8_deg}
% \end{figure}

\begin{figure}
    \centering
    \includegraphics[width=\linewidth]{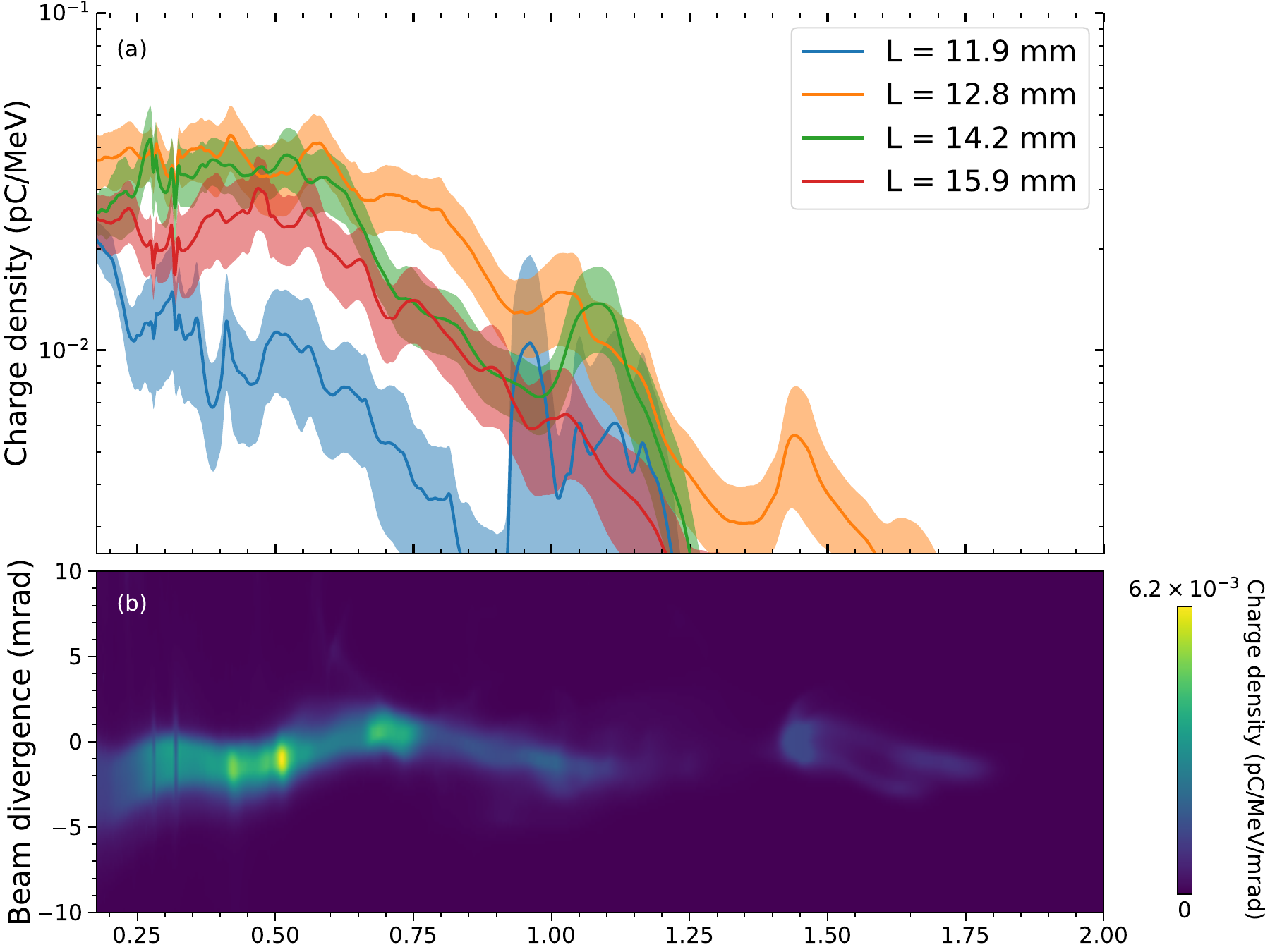}
    \caption{Spectra obtained with a tilting angle of 8°. (a) Mean spectra obtained  for different acceleration length. The mean is computed over 10 shots, without excluding any, for each pressure and the shaded area corresponds to the standard deviation over the same 10 shots. (b) Angularly resolved spectra of a selected shot, with a linear colorbar.}
    \label{fig:20240606_8_deg}
\end{figure}

%For an operating pressure of 26~bar and a tilting angle of 8°, we were able to scan the target length to scan the dephasing process. 

We then tested different configurations of tilted nozzles to mitigate dephasing and found an optimum at an angle of 8° and a backing pressure of 26~bar. The corresponding density profile is shown in Supplementary Fig.~S4. In this configuration, the initial plasma density without guiding is on the order of $n_0 = 8.6\times10^{18} ~\text{cm}^{-3}$, with a density gradient of $0.03n_0/\text{mm}$. The effective plasma length was scanned by translating the nozzle underneath the blade.
Mean spectra for each target length are shown in Fig.~\ref{fig:20240606_8_deg}a. An optimum is observed  at $L_\text{{target}}=12.8$~mm, where the beam reaches a cut-off energy of $1.6\pm0.1$~GeV, corresponding to an increase in electron energy by a factor of $\approx1.4$ compared to the reference case, thus demonstrating the effectiveness of plasma tapering in mitigating dephasing. At longer target lengths, the cut-off energy decreases noticeably, down to %$1.3\pm0.03$~GeV for $L_{\text{target}}=14.2$~mm and
$1.2\pm0.1$~GeV for $L_{\text{target}}=15.9$~mm. %Averaging over 10 shots for each target length leads to a maximum cut-off energy of $1.3\pm0.52$~GeV for $L_{target}=12.8$~mm.As we increase the target length, we observe a slight decrease of the energy to $1.1\pm0.33$~GeV and $1.0\pm0.46$~GeV for $L_{target}=14.2$ and 15.9~mm. 

%The charge above 200 MeV initially increases with the target length due to continuous injection, rising from $7.7\pm3.6$~pC for $L_{\text{target}}=11.9$~mm,  to $28\pm6.6$~pC for $L_{\text{target}}=12.8$~mm. For longer target lengths, the injected charge decreases down to %$22\pm4.7$~pC for $L_{\text{target}}=14.2$~mm, and to
%$16\pm4.2$~pC for $L_{\text{target}}=15.9$~mm, consistent with laser depletion and dephasing. 
The charge above 1~GeV initially increases with the target length, increasing from $1.3\pm1$~pC for $L_{\text{target}}=11.9$~mm to $4.2\pm1.4$~pC for target lengths of 12.8~mm, before decreasing to %$2.6\pm0.8$~pC for $L_{\text{target}}=14.2$~mm, 
$1.6\pm0.6$~pC for $L_{\text{target}}=15.9$~mm
The performance achieved with this density gradient is further illustrated in Fig.~\ref{fig:20240606_8_deg}b, which shows an angularly resolved spectra for a selected shot, with some electrons reaching energies exceeding 1.8~GeV. In (b), the beam charges above 1 and 1.5 GeV are 9.4~pC and 3~pC, respectively.

\begin{figure}
    \centering
    \includegraphics[width=\linewidth]{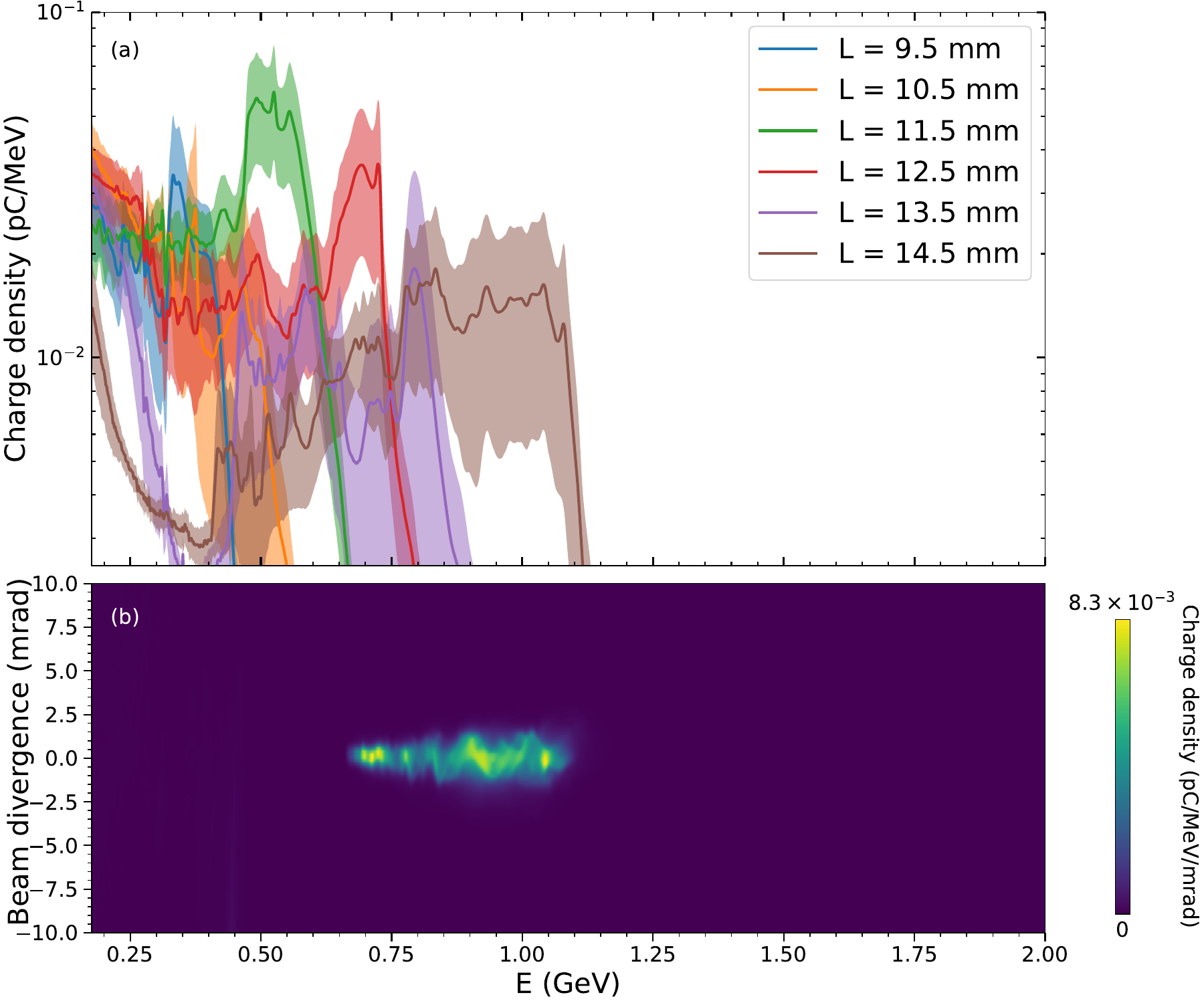}
    \caption{Spectra obtained with a tilting angle of 12°. (a) Mean spectra for different acceleration length. The mean is computed over 10 shots, without excluding any, for each target length and the shaded area corresponds to the standard deviation over the same 10 shots. (b) Angularly resolved spectra of a selected shot, with a linear colorbar.}
    \label{fig:20240610_12_deg_30b}
\end{figure}

%For an operating pressure of 30~bar, and a tilting angle of 12°, 
The model presented in Sec.~\ref{sec:theo} is too simplified to quantitatively predict the optimal density gradient under our experimental conditions. We therefore experimentally tested several other density gradients. Figure~\ref{fig:20240610_12_deg_30b} shows the average spectra obtained for an initial density (before channel generation) of $n_0 = 1.1\times10^{19}\text{cm}^{-3}$ and a nozzle inclination of 12°, resulting in a density gradient of $0.06n_0/\text{mm}$ (see the corresponding density profile in Supplementary Fig.~S5). Compared to the 8° inclination case, we observe an overall decrease in energy, although the electron energy still increases with target length. This reduced energy suggests a delayed injection, which limits the effective acceleration length. The delayed injection is likely caused by an overly steep density gradient that produces a superluminal phase velocity at the beginning of propagation, thereby hindering early injection. %In addition, the higher average density after injection leads to faster dephasing, thus reducing the final energy.  
This scenario would correspond to the one illustrated in Fig.~\ref{fig:comp_gain} for $\alpha=1.5$. A second effect attributed to an excessively delayed injection is the shorter total injection length, which leads to an energy distribution peaked at high energy, as shown in Fig.~\ref{fig:20240610_12_deg_30b}b.

Quantitatively, the cut-off energy remains below 1.15~GeV for all target lengths. The total charge above 200~MeV is relatively stable for $L_\text{target} > 10$~mm, reaching approximately 15~pC, while the charge above 1~GeV reaches a maximum of $1.6\pm0.9$~pC for $L_\text{target} = 14.5$~mm, which was the maximum target length we could reach in this configuration. The absence of evidence for dephasing, particularly the steady increase in cut-off energy with target length led us to increase the plasma density. We found that a 50\% increase in density resulted to clear evidence of dephasing around $L_\text{target} \sim 12.5$~mm. However, this was accompanied by lower electron energies, reduced charge above 1~GeV, and more irregular energy-angle distributions. These results suggest that the initial plasma density was close to optimal for a 12° inclination. They also confirm the superior performance of the 8° configuration, which was further supported by tests at a 3° inclination. In this case, the results were similar to those obtained at 8°, but with the acceleration of slightly lower-energy electrons.

\section{Numerical simulation}

To support our interpretation of the experimental result, we ran simulations using the pseudo-spectral quasi-cylindrical particle-in-cell code FBPIC \cite{lehe2016spectral}. We assumed a perfect gaussian beam with an energy of 1~J, a radius at FWHM of $25~\mu$m, and a gaussian temporal enveloppe of 25~fs, corresponding to $a_0\approx 1.6$. To investigate different interaction regimes, we tested various initial plasma densities and density gradients, which enabled us to distinguish between different regimes of interaction.  For simplicity, the plasma was assumed to pre-ionised with a pre-formed plasma channel. % whose radius decreased along the propagation axis to remain matched to the laser spot size and the local plasma wavelength $\lambda_p\propto n_e(z)**{-1/2}$ 

A key challenge in combining plasma tapering with guiding is ensuring that the waist of a matched laser evolves as the radius of the plasma cavity, which scales with the plasma wavelength $\lambda_p \propto n_e^{-1/2}$. To maintain efficient guiding and avoid significant laser leakage, the waveguide diameter should follow the same evolution. In the case of a linear density gradient, this implies that the waveguide diameter, and thus the waist of the HOFI laser, should scale as $z^{-1/2}$. Such a decreasing waist naturally arises from quasi-Bessel beams, like those produced at the focus of an axiparabola. In particular, an axiparabola designed for a constant-intensity focal line generates a focal spot whose diameter scales precisely as $z^{-1/2}$, potentially enabling perfect matching, provided the proportionality factor is correct. This factor is primarily determined by the axiparabola’s numerical aperture and focal depth~\cite{Oubrerie_2022}, which should thus be specifically chosen to match a given density gradient.

In our experiment, we used a single, constant-encircled-energy axiparabola and varied the target density gradient. As a result, the waveguide diameter did not perfectly follow the evolution of $\lambda_p$, which may have led to increased laser leakage during acceleration. However, to simplify the analysis and limit the number of parameters, we initially neglect this effect and assume in the simulations an ideal parabolic channel that follows the matched guiding of the drive laser with a constant matched spot size $w_0$ of the form %that the radius of the preformed plasma channel decreases along the propagation axis so that it continuously matches the local plasma wavelength.
\begin{equation}
    n_e(r,z) = n_0(z) + \frac{1}{\pi r_e w_0^4}\left(\frac{r}{w_0}\right)^2,
    \label{eq:radial_channel}
\end{equation}
with $r_e$ the electron classical radius.

% \begin{figure}
%     \centering
%     \includegraphics[width=\linewidth]{Figures/case_1_5_matched.pdf}
%     \caption{PIC simulation for an initial density of $n_0=6.2\times10^{17}~\text{cm}^{-3}$ and a density gradient of $0.06n_0/\text{mm}$.(a-c) Reference case without gradient. (d-f) Gradient case. (a,d) Energy evolution of test particles selected from the high-energy tail. The particles are selected by looking for the peak in the spectrum at highest energy, with a peak charge density above 2.5~fC/MeV. The spectral dispersion of the bunch selected is equal to the full-width at half-maximum of the peak. (b,e) On-axis longitudinal electric field $E_z$; the blue line indicates the position of the peak transverse field $E_x$ (i.e., the laser pulse maximum), while the red lines mark the positions of the test electrons. (c,f) Evolution of the normalized laser amplitude $a_0$ along the propagation axis.}
%     \label{fig:case_1_5_matched}
% \end{figure}

\begin{figure*}[t!]
    \centering
    \includegraphics[width=\textwidth]{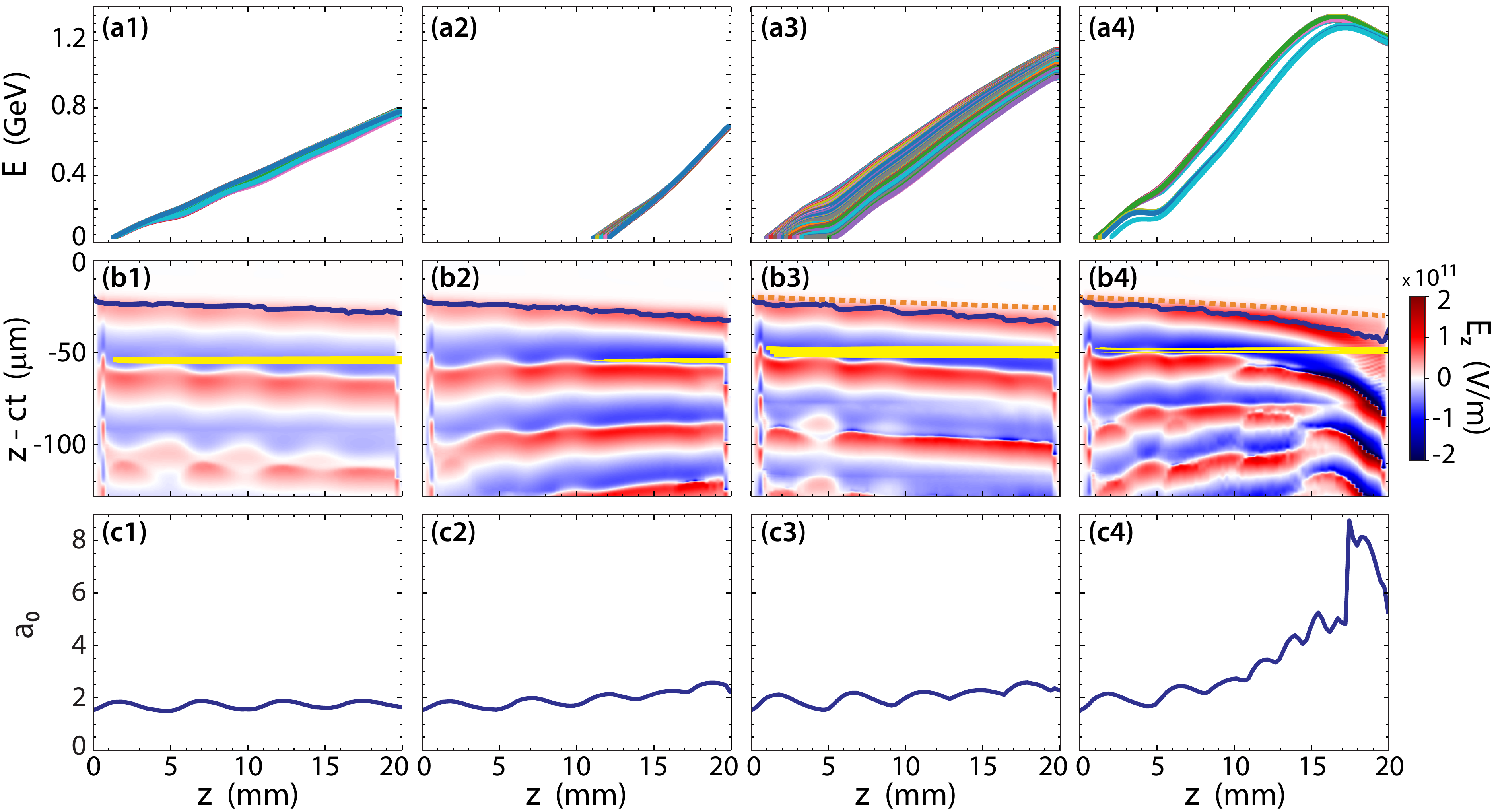}
    \caption{Result of PIC simulations. Columns~1 and~2 are for an initial density of $n_0=6.2\times10^{17}~\text{cm}^{-3}$. Columns~3 and ~4 are for an initial density of $n_0=1.0\times10^{18}~\text{cm}^{-3}$. Columns~1 and ~3 are the reference case without gradient, for these two densities. Column~2 is for a density gradient of $0.06n_0/\text{mm}$, and column~4 is for a density gradient of $0.08n_0/\text{mm}$. (a) Energy evolution of test particles selected from the high-energy tail. The particles are selected by looking for the peak in the spectrum at highest energy, with a peak charge density above 2.5~fC/MeV. (b) On-axis longitudinal electric field $E_z$; the blue line indicates the position of the peak transverse field $E_x$ (i.e., the laser pulse maximum), while the yellow lines mark the positions of the test electrons. (c) Evolution of the normalized laser amplitude $a_0$ along the propagation axis.}
    \label{fig:PIC_full}
\end{figure*}

Figure~\ref{fig:PIC_full}(a-c 1-2) reports on the case of an initial density $n_0=6.2\times10^{17}~\text{cm}^{-3}$ without gradient (a-c1), or with a
gradient $\delta_n= 0.06n_0/\text{mm}$ (a-c2).  The panels (a1) and (a2) show the evolution of the energy of a few electrons from the high-energy tail. In both cases, the electrons reach a final energy of approximately 750–800~MeV. % after 20~mm of propagation. 
However, in the gradient case (a2), this occurs over nearly half the distance, due to delayed injection and a stronger average accelerating field. %, which stems from the increasing plasma density.
The delayed injection is caused by an initially superluminal wakefield (corresponding to the case $\alpha >1$ in Sec.~\ref{sec:theo}).
%However, in the gradient case (d), this energy is reached over a distance nearly half as long . This results from delayed injection and a stronger average accelerating field due to the increasing plasma density. The delayed injection stems from an initially superluminal wakefield (corresponding to the case $\alpha >1$ in Sec.~\ref{sec:theo}).
This is evidenced in Fig.~\ref{fig:PIC_full}(b2), which shows the amplitude of the wakefield $E_z$ as a function of the co-moving coordinate $\xi =z-c t$ and the longitudinal position $z$. In this coordinate system, a luminal wake propagates horizontally, while a superluminal one appears with a positive slope.  As shown, the wakefield is superluminal up to $z\approx 10$~ mm at which point injection occurs. This scenario is thus close to the one depicted by the red curves in Fig.~\ref{fig:comp_gain}.

We also observe that the electron beam (indicated by the thick yellow lines in Fig.~\ref{fig:PIC_full}(b2)) remains in the first quarter of the wakefield, far from the dephasing region. This suggests that longer acceleration lengths would have led to higher final energies.  According to Lu’s model~\cite{PhysRevSTAB.10.061301}, the dephasing length for the constant density case is estimated as $L_{D0} \approx 33~\text{mm}$, which is consistent with the absence of dephasing over 20~mm. Moreover, this estimate yields $\alpha_{\text{theo}} = \delta_n L_{D0} / n_0 \approx 2$ which is also consistent with the wakefield being initially superluminal.

Panels (c1) and (c2) in Fig.~\ref{fig:PIC_full} show laser amplitude oscillations along the propagation, pointing to imperfect matching. In addition to this,  the gradient case exhibits an increase in amplitude, revealing self-focusing. This effect gradually increases the non-linear plasma wavelength, causing the back of the plasma cavity to slow down. This effect, not accounted for in the simplified model of Sec.~\ref{sec:theo}, tends to enhance dephasing and shorten the effective dephasing length, which in turn lowers the effective $\alpha$.%The position of the laser field maximum is overlaid as a blue line in the second column, highlighting that the simple model in Sec.\ref{sec:theo} does not account for laser etching effects. This etching shifts the laser peak position, further modifying the 

We draw two conclusions from this initial simulation: to increase the final energy, the plasma density should be increased to better exploit the target length, and the effective $\alpha$ factor should be lowered to enable earlier injection. Figure~\ref{fig:PIC_full}(a-c4) presents the results obtained for $n_0 = 1 \times 10^{18}$~c$^{-3}$ and $\delta_n = 0.08n_0/\text{mm}$, which gives $L_{D0} \approx 15$~mm and $\alpha_{\text{theo}} \approx 1.2$. According to Sec.\ref{sec:theo}, such a value of $\alpha$ may still inhibit injection in the early stage of propagation. However, we previously saw that the effective $\alpha$ can be significantly reduced due to self-focusing. Here, Fig.~\ref{fig:PIC_full}(c4) shows that self-focusing is quite strong in the gradient case, with the laser amplitude nearly tripling over 16~mm.
Other effects also contribute to modifying the wakefield velocity by altering the laser group velocity. This is illustrated in Fig.~\ref{fig:PIC_full}(b3-4), where the orange dashed line shows the trajectory of an entity propagating at the group velocity $v_g(z)$. While the front of the pulse follows this trajectory, the maximum of the laser pulse (blue curve) moves significantly slower due to self-phase modulation, dispersion, and laser etching, which in turn further slows down the wakefield.
The combination of these effects explains why efficient injection occurs just after the shock region, % as shown in Fig.\ref{fig:case_2_5_matched}(d,e), 
even though $\alpha_{\text{theo}} > 1$.

Figure~\ref{fig:PIC_full}(a,b 3) shows that in the constant-density case, the beam is close to dephasing after approximately $19$~mm of acceleration, in qualitative agreement with the estimate of $L_{D0}$. In contrast, in the density gradient case, panel (b4) shows that the beam hardly dephases over the first $\approx 10$~mm, before dephasing rapidly, leading to a total acceleration length of about $16$~mm. The scenario is therefore close to the case $\alpha = 1$ in Fig.~\ref{fig:comp_gain} (green curves), except for the faster dephasing due to the strong increase in laser intensity toward the end of the plasma. 
The efficient tapering increases the electron beam energy by almost 20\% compared to the constant-density case, resulting in a final energy approaching $1.4$~GeV, close to the experimentally measured value.

\begin{figure}
    \centering
    \includegraphics[width=\linewidth]{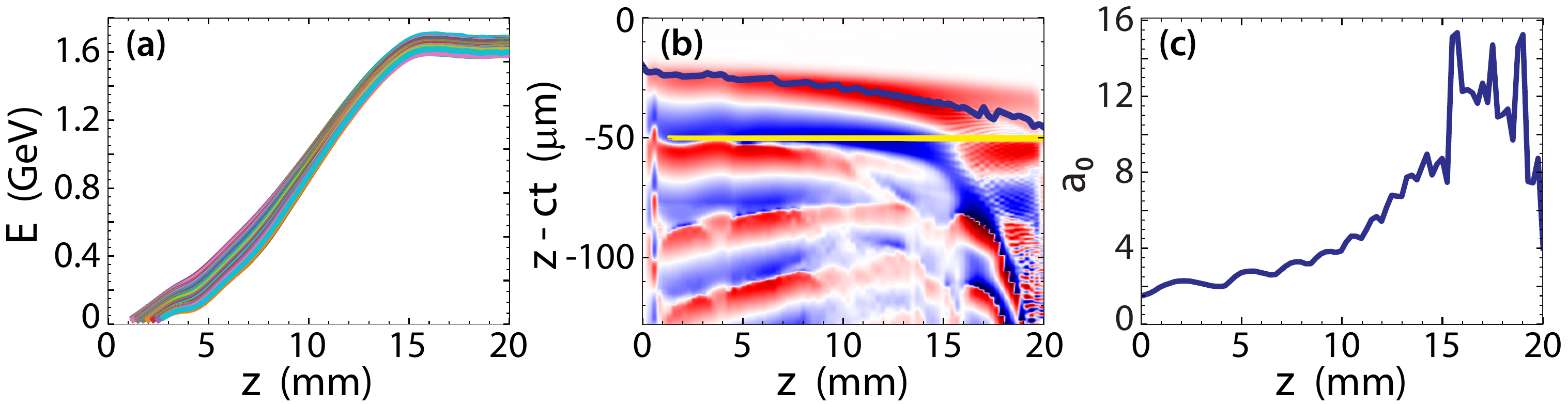}
    \caption{PIC simulation for an initial density of $n_0=1.0\times10^{18}~\text{cm}^{-3}$ and a density gradient of $0.08n_0/\text{mm}$, with a channel transverse size evolving as the spot size of the channel forming beam. (a) Energy evolution of test particles selected from the high-energy tail. The particles are selected by looking for the peak in the spectrum at highest energy, with a peak charge density above 2.5~fC/MeV. The spectral dispersion of the bunch selected is equal to the full-width at half-maximum of the peak. (b) On-axis longitudinal electric field $E_z$; the blue line indicates the position of the peak transverse field $E_x$ (i.e., the laser pulse maximum), while the red lines mark the positions of the test electrons. (c) Evolution of the normalized laser amplitude $a_0$ along the propagation axis.}
    \label{fig:PIC_not_matched}
\end{figure}

However, an important physical effect is still missing in this simulation: the influence of the  transverse size of the channel forming beam on the waveguide. The transverse size of the channel-forming beam depends on the local numerical aperture~\cite{Oubrerie_2022}, which results, in our case, in a decreasing transverse size along the propagation To address this, we consider in Fig.~\ref{fig:PIC_not_matched} a non-ideal waveguide. More precisely, we model a waveguide following Eq.~\ref{eq:radial_channel}, but by replacing $w_0$ with $w_m(z) = C(z)\times w_0$, with $w_0$ the size of the drive laser, and $C(z)$ a numerical factor scaling as the transverse size of the channel forming beam. The reduction of the matched spot-size during the propagation promotes self-focusing, as shown in Fig.~\ref{fig:PIC_not_matched}(c) by the rapid increase of the laser amplitude. This increase in the laser amplitude results in a cavity becoming fully evacuated more efficiently, which has the effect of increasing the amplitude of the accelerating field. This higher amplitude of the accelerating field results in a higher energy gain, about 1.6~GeV, as shown in Fig.~\ref{fig:PIC_not_matched}(a) while also increasing the dephasing due to the increase of laser intensity. Other effects, such as the truncation of the channel, or a non-parabolic radial profile could also be taken into account, but are beyond the scope of this study.

%Second row of Fig.~\ref{fig:case_2_5_matched} shows dephasing occurring as we increase the initial density and the density gradient [see code].
%We can observe that the deceleration of the wake at the end of the propagation is associated with an increase of $a_0$, due to a combination of focusing along the propagation because of the decreasing channel radius, and relativistic self-focusing occurring as the density decreases. This increase of $a_0$ results in an increase of the size of the cavity, which has the effect of slowing down the wake. 

\section{Conclusion}
We have experimentally demonstrated that combining plasma tapering with optical guiding enables effective mitigation of dephasing in a laser-plasma accelerator. By introducing a linear density gradient in a guided plasma, we achieved a 40\% increase in electron beam energy, reaching up to 1.8~GeV with a Joule-class laser. These results are supported by particle-in-cell simulations, which emphasize the crucial role of nonlinear laser propagation and laser-plasma matching throughout the acceleration process.

 Looking ahead, further energy gains could be achieved by employing more complex, non-linear density profiles designed to lock the wakefield velocity to that of the electron beam. Such profiles could be engineered using techniques such as laser machining of gas targets~\cite{PhysRevLett.94.115003}, or a more precise control of the target design~\cite{10.1063/5.0250698}. Another promising avenue is the implementation of plasma tapering combined with guiding on petawatt-class laser systems, which may enable the production of electron beams exceeding 10~GeV. These developments would mark significant progress toward high-energy laser-plasma accelerators for future light sources and high-energy physics applications.

\bibliographystyle{apsrev4-2}
\bibliography{sample}

\end{document}